\documentclass[11pt,preprint]{aastex}
\usepackage{lscape}
\usepackage{epstopdf}
\usepackage{color}
\usepackage{epsfig}
\usepackage{comment}
\usepackage{amsmath}
\usepackage{graphicx}
\usepackage{lineno}
\shorttitle{X-ray Emission and Radio Emission from the Jets and Lobes of the Spiderweb Radio Galaxy}

\shortauthors{Carilli, Anderson, Tozzi, et al.}
 
\begin{document}

\title{X-ray Emission from the Jets and Lobes of the Spiderweb}

\author{
Christopher L. Carilli\altaffilmark{1},
Craig S. Anderson\altaffilmark{2},
Paolo Tozzi\altaffilmark{3},
Maurilio Pannella\altaffilmark{4},
Tracy Clarke\altaffilmark{5},
L. Pentericci\altaffilmark{14},
Ang Liu\altaffilmark{3},
Tony Mroczkowski\altaffilmark{6},
G.K. Miley\altaffilmark{7},
H.J. Rottgering\altaffilmark{7},
S. Borgani\altaffilmark{8, 9, 10, 11},
Colin Norman\altaffilmark{12,13},
A. Saro\altaffilmark{8,9,10,11},
M. Nonino\altaffilmark{9},
L. Di Mascolo\altaffilmark{8}
}

\altaffiltext{1}{National Radio Astronomy Observatory, P. O. Box 0,Socorro, NM 87801, USA, ccarilli@nrao.edu, ORCID: 0000-0001-6647-3861}

\altaffiltext{2}{Jansky Fellow of the National Radio Astronomy Observatory, P. O. Box 0,Socorro, NM 87801, USA}

\altaffiltext{3}{INAF - Osservatorio Astrofisico di Arcetri, Largo E. Fermi 5, 50127, Firenze, Italy}  

\altaffiltext{4}{Astronomy Unit, Department of Physics, University of Trieste, via
Tiepolo 11, I-34131 Trieste, Italy}                       

\altaffiltext{5}{Naval Research Laboratory, Code 7213, 4555 Overlook Ave SW, Washington, DC 20375, USA}

\altaffiltext{6}{Max Planck Institute for Extraterrestrial Physics, Giessenbachstrasse 1, 85748 Garching, Germany}

\altaffiltext{6}{European Southern Observatory, Karl-Schwarzschild-Str. 2, D-85748 Garching b. Munchen, Germany}

\altaffiltext{7}{Leiden Observatory, Leiden University, P.O.Box 9513, NL-2300 RA, Leiden, The Netherlands}

\altaffiltext{8}{Dipartimento di Fisica dell’ Università di Trieste, Sezione di Astronomia, via Tiepolo 11, I-34131 Trieste, Italy}

\altaffiltext{9}{INAF-Osservatorio Astronomico di Trieste, Via G. B. Tiepolo 11, I-34131 Trieste, Italy}

\altaffiltext{10}{INFN, Sezione di Trieste, Via Valerio 2, I-34127 Trieste TS, Italy}

\altaffiltext{11}{IFPU, Institute for Fundamental Physics of the Universe, via Beirut 2, 34151 Trieste, Italy}

\altaffiltext{12}{ Space Telescope Science Institute, 3700 San Martin Dr., Baltimore, MD 21210, USA}

\altaffiltext{13}{Johns Hopkins University, 3400 N. Charles Street, Baltimore, MD 212}

\altaffiltext{14}{INAF-OAR, Via Frascati 33, Monte Porzio Catone (RM), Italy}

\begin{abstract}

Deep Chandra and VLA imaging reveals a clear correlation between X-ray and radio emission on scales $\sim 100$~kpc in the Spiderweb radio galaxy at z=2.16. The X-ray emission associated with the extended radio source is likely dominated by inverse Compton up-scattering of cosmic microwave background photons by the radio emitting relativistic electrons. For regions dominated by high surface brightness emission, such as hot spots and jet knots, the implied magnetic fields are $\sim 50~\mu$G to $70~\mu$G. The non-thermal pressure is these brighter regions is then $\sim 9\times 10^{-10}$ dyne cm$^{-2}$, or three times larger than the non-thermal pressure derived assuming minimum energy conditions, and an order of magnitude larger than the thermal pressure in the ambient cluster medium. Assuming ram pressure confinement implies an average advance speed for the radio source of $\sim 2400$ km s$^{-1}$, and a source age of $\sim 3\times 10^7$ years. Considering the lower surface brightness, diffuse radio emitting regions, we identify an evacuated cavity in the Ly$\alpha$ emission coincident with the tail of the eastern radio lobe. Making reasonable assumptions for the radio spectrum, we find that the relativistic electrons and fields in the lobe are plausibly in pressure equilibrium with the thermal gas, and close to a minimum energy configuration. The radio morphology suggests that the Spiderweb is a high-$z$ example of the rare class of hybrid morphology radio sources (or HyMoRS), which we attribute to interaction with the asymmetric gaseous environment indicated by the Ly$\alpha$ emission.  

\end{abstract}


\section{Introduction}

It is well documented that powerful radio AGN act as beacons to massive galaxy formation in protoclusters in the early Universe  \citep{md, carilli01}. The powerful radio emitting jets are driven by mass accretion onto supermassive black holes, and are thought to play a number of important roles in galaxy formation. Radio-mode feedback, meaning heating of the protocluster medium around the host galaxy by the radio jets on tens to hundreds of kpc scales, is often invoked to suppress late-time gas accretion, thereby limiting high mass galaxy growth \citep{fabian12, mcnamara05}. Conversely, jet-induced star-formation, induced by compression of dense gas in the host galaxies by shocks driven by the advancing radio source, may enhance star formation locally in some systems \citep{gaibler}. The radio jets may also remove angular momentum from the AGN accretion disk, thereby enhancing accretion rates and early SMBH formation \citep{bp82}.  These same jets can also transport magnetic fields and possibly heavy elements to large scales in the protocluster \citep{ct02}. 

Among the best studied powerful radio galaxies in a dense environment at large look back time is the Spiderweb radio galaxy, J1140-2629, at $z = 2.16$ \citep{pentericci97, pentericci98, kurk00, miley06, md}, with high resolution imaging and spectroscopy at optical through radio wavelengths. The radio source is characterized by two jets oriented roughly east-west, extending about $7''$ ($\sim 60$ kpc) from the AGN. The protocluster shows a dense distribution of galaxies extending to radii $> 100$ kpc, as well as a Ly$\alpha$ halo on a similar scale, with a major axis along the jet axis \citep{miley06}. Many of these galaxies have been detected in CO and thermal emission from warm dust, including the host galaxy of the radio source itself \citep{gullberg, dannerbauer14, jin21}. There also appears to be a diffuse CO emitting component on $\sim 100$ kpc scale \citep{emonts}.

A 28 ksec (on-source) observation with {\sl Chandra} revealed a dominant X-ray AGN, plus a faint extension of the X-ray emission along the radio jet \citep{carilli02}. However, a clean separation of the faint extended emission from the AGN was difficult with the limited exposure, and physical conclusions speculative. To date, a detailed understanding of the physical conditions (pressures, magnetic fields), in the radio source and the ambient medium in the Spiderweb complex remains limited. 

We have embarked on a deep study of the Spiderweb galaxy using {\sl Chandra} and the Jansky Very Large Array. We have obtained a 700~ksec {\sl Chandra} exposure, as well as broad band radio observations with arcsecond to sub-arcsecond resolution from 340 MHz to 36 GHz, and low spatial resolution observations with the ALMA Small Baseline Array (Saro et al. in prep). In this paper, we present the extended X-ray emission associated with the radio jets and lobes on $\sim 100$~kpc scales. We derive magnetic fields and pressures in the source under the standard assumptions of inverse Compton up-scattering of CMB photons by the relativistic synchrotron emitting electrons in the outer regions of the source $> 20$~kpc from the core. A future analysis will consider contributions of other photon fields over the full source \citep{sobolewska}. In parallel papers we explore the polarized radio emission and, in particular, study the magnetic field orientation in the radio source, and the extreme rotation measures indicative of a magnetized ambient thermal gas \citep{anderson21}, the proto-cluster AGN population based on the deep X-ray exposure \citep{tozzi21}, and the nuclear X-ray source and diffuse emission away from the radio jets \citep{tozzi22}.

\section{Observations and data reduction}

The Spiderweb Galaxy was the target of a {\sl Chandra} Large Program observation with ACIS-S (AO20, PI P. Tozzi). Data reduction is performed starting from the level-1 event files with {\tt CIAO 4.13}, with the latest release of the {\sl Chandra} Calibration Database at the time of writing {\tt (CALDB 4.9.3)}. The details of data reduction, including removal of spurious events and residual cosmic rays, and high-background filtering, are described in \citet{tozzi21}.   The total effective exposure time is 715~ks.

We obtained deep VLA observations in September, 2019 at X band (8 GHz to 12 GHz) at $0.4''\times 0.2''$ resolution (major axis oriented north-south, in all cases), and S band (2 GHz to 4 GHz), at $1.3''\times 0.6''$ resolution.  We use '3 GHz' to designate the full S band, and '10 GHz' to designate the full X band. A total of 8 hours was observed in each band. 

Standard pipeline calibration was employed at the VLA\footnote{\url{https://science.nrao.edu/facilities/vla/data-processing/pipeline}}, 
using 3C 286 as the absolute flux density scale calibrator. The calibrated data were then threshold flagged, and self-calibrated using both AIPS and CASA. The final imaging process employed the CLEAN algorithm, with Briggs weighting and a robust factor of zero \citep{briggs}. Multiscale clean was employed, cleaning down to 2$\sigma$ residuals. Bandwidth synthesis was also employed, across four 0.5 GHz bands spanning the 3~GHz band, and eight 1 GHz bands across the 8~GHz band. In each case, the images were restored with CLEAN Gaussian beams corresponding to the resolution at mid-band. The final images were summed to produce a single image for each band, while the sub-band images were also analyzed for spectral index across a given band. The resulting rms noise on the full band images was 3$\mu$Jy beam$^{-1}$ at 10 GHz, and 5$\mu$Jy beam$^{-1}$ at 3 GHz.

As part of the VLITE system \citep{vlite}, commensal observations at 340 MHz (P-band) with a resolution of $8.0'' \times 4.3''$ (major axis north-south), were obtained in parallel with our higher frequency program. The VLITE 340 MHz data were processed using a dedicated flagging and calibration pipeline developed within the Obit \citep{Obit} and AIPS \citep{AIPS} data reduction packages. The pipeline uses automated flagging routines to remove radio frequency interference followed by standard delay, gain, and bandpass calibrations. The calibrated VLITE data were imaged using the wide-band imaging algorithm MFimage in Obit to map a region of radius 1$^\circ$ from the pointing center with small facets. Two rounds of phase-only self calibration were applied to the data. While all four days of observations were recorded with VLITE, the period of time in early September 2019 was known to be significantly impacted by ionospheric fluctuations, thus we concentrate on the lowest noise day (rms noise 640 $\mu$Jy beam$^{-1}$, Sept.\ 13, 2019) for our analysis herein. 

\begin{figure}
\includegraphics[trim=0.5in 2in 0in 0in, clip, width=1.3\linewidth]{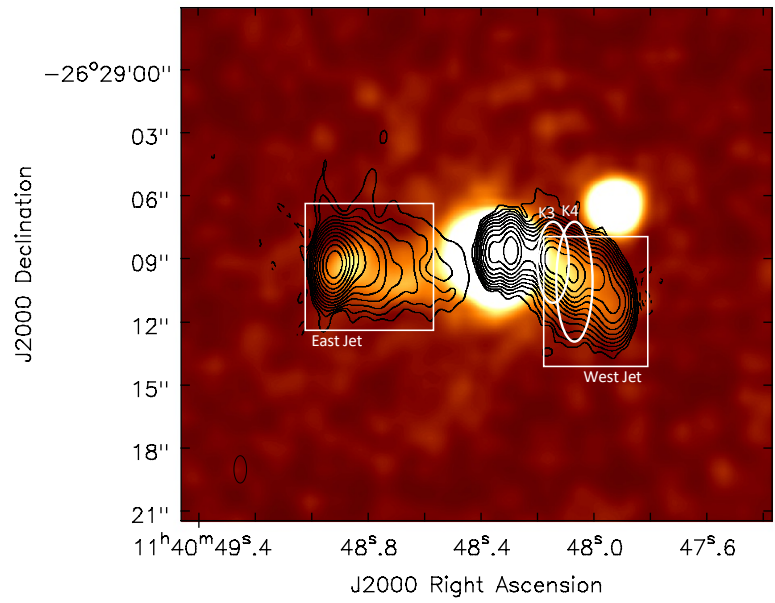}
\caption{Same X-ray total counts color image and 3 GHz contours, with boxes and ellipses indicating the four regions for the inverse Compton and Synchrotron analysis in Section~\ref{sec:mech}.
}
\label{fig:regions}
\end{figure}

When comparing the {\sl Chandra} X-ray and VLA radio images, we employ the standard astrometry for the {\sl Chandra} observatory, as well as the astrometry set by the VLA calibrators, using J2000 coordinates. A check of the relative astrometry is given by the position of the active nucleus of the Spiderweb, for which we find agreement between the radio and X-ray to within $0.1''$. The X-ray and radio alignments we consider herein are for structures on significantly large scales, $\sim 1"$ or greater. In particular, the quantitative spectra considered average over areas with few arcsec scale (see Figure~\ref{fig:regions}).

We also employ an archival image of the Ly$\alpha$ emission from the Spiderweb. Narrow band imaging was carried out on April 1999 with ESO VLT Antu (UT1) using FORS1 in imaging mode (ESO programme 63.O-0477(A), P.I. Miley \citep{kurk00}. The narrow band filter has a central wavelength of 381.4 nm and a FWHM of 6.5~nm. The diffuse emission in this filter is dominated by the rest frame Ly$\alpha$ emission from the Spiderweb, although there remains some continuum emission in the filter as well. Details of these VLT observations and results will be presented in \citet{nonino}. Herein, we present a morphological comparison of the Ly$\alpha$ and radio emission.

\section{Results}

\subsection{Total Intensity Images}

\begin{figure}
\includegraphics[trim=-1in 4.7in 0in 0.5in, clip, width=\linewidth]{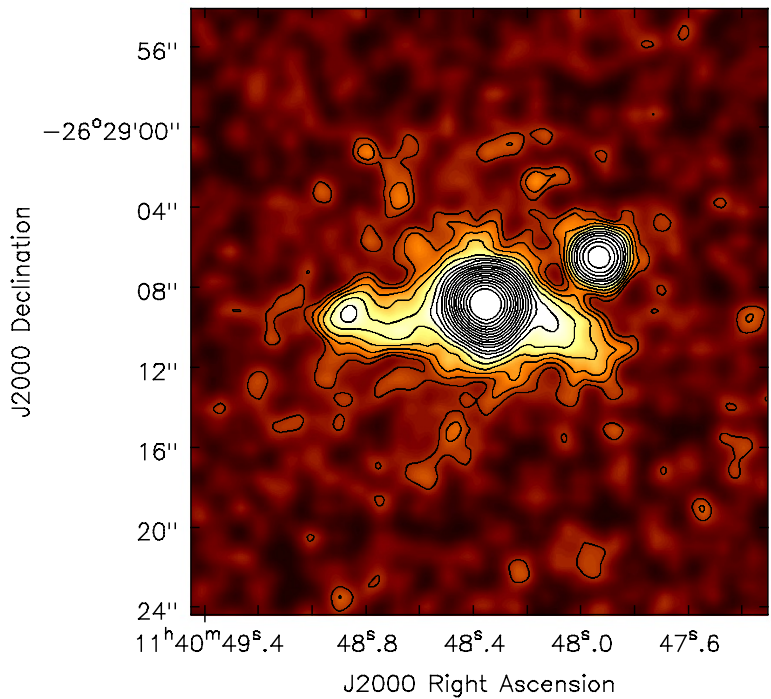}
\caption{Contour and color scale image of the $z = 2.16$ powerful radio galaxy,  J1140-2629 (the 'Spiderweb galaxy'), from {\sl Chandra} Observatory total X-ray emission (0.5 keV to 7 keV), convolved with a $1''$ Gaussian. This is the total counts image. Further processing, including AGN subtraction, will be presented in \citet{tozzi21}. The contour levels are a geometric progression in the square root two, starting at 0.8 counts beam$^{-1}$. 
}
\label{fig:Xray}
\end{figure}

\begin{figure}
\includegraphics[trim=-1in 5in 0in 0.5in, clip, width=\linewidth]{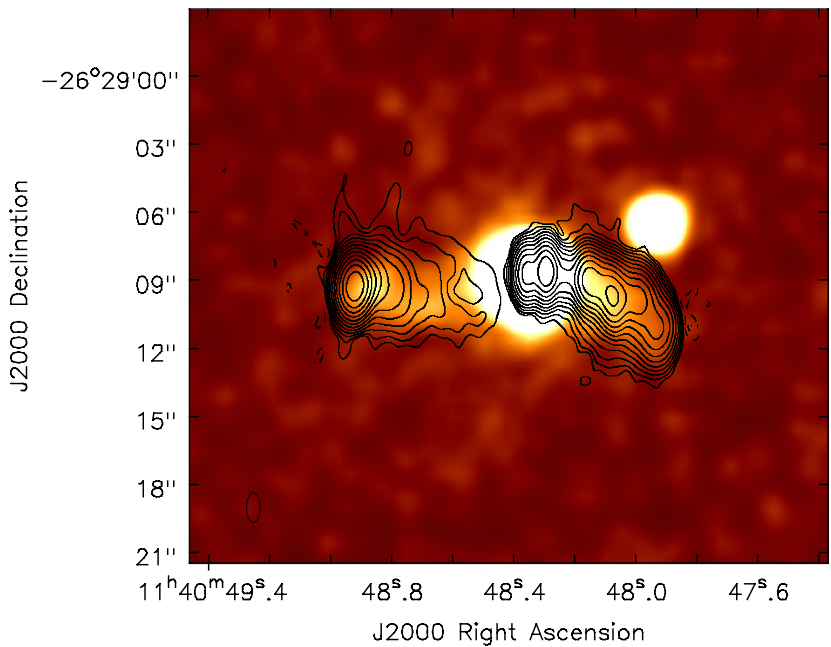}
\caption{Contour image of the 2 GHz to 4 GHz VLA image of the Spiderweb galaxy, at a resolution of $1.3''\times 0.6''$, major axis north-south. The contour levels are a geometric progression in a factor two, starting at 20$\mu$Jy beam$^{-1}$. Negative contours are dashed. The color scale is the same  {\sl Chandra} Observatory total X-ray emission (0.5 keV to 7 keV), as shown in Figure~\ref{fig:Xray}. 
}
\label{fig:XraySband}
\end{figure}

Figure~\ref{fig:Xray} shows the total X-ray emission from 0.5 to 7 keV from the {\sl Chandra} Observatory as both contour and color scale. The X-ray counts image has been convolved with a $1''$ Gaussian. The range 7~keV to 10~keV is not used for imaging since the signal is negligible and the background very high.  However, the X-ray fluxes we derive refer to the range 0.5 keV to 10 keV (or separately 0.5 keV to 2 keV and 2 keV to 10 keV), since they are obtained from the full-band spectral analysis. Figure~\ref{fig:XraySband} shows the same X-ray image as color scale, with the VLA 3 GHz contour image at $1.3''\times 0.6''$ resolution (major axis north-south).

The net counts (background subtracted) for the full source in the 0.5-7 keV X-ray band is 10400, which is dominated by emission from the AGN, which has $\sim 9600$ counts. The net 0.5~keV to 7~keV X-ray counts for the western jet, outside of $2''$ radius from the core, is 200, while that for the eastern jet is 272. There is a point X-ray source $5''$ northwest of the AGN which appears to be associated with an optical galaxy in the protocluster at a large velocity offset from the main Spiderweb host galaxy.  Likewise, there appears to be faint, diffuse emission north and south of the AGN, possibly indicating thermal emission from a hot ambient medium on tens of kpc-scales. We consider the AGN and other X-ray point sources, and thermal emission, in parallel papers \citep{tozzi21,tozzi22}. The focus of this paper is the extended X-ray emission associated with the radio jets. 

The rms noise on the 3 GHz image is $5.5~\mu$Jy beam$^{-1}$. The radio emission to the west shows a prominent jet, defined by a series of hot spots and more diffuse emission, extending at least $7''$ ($\sim 60$ kpc), from the AGN, and becoming fainter with distance from the AGN. The emission to the east shows a bright radio hot spot at the source extremity, about $7''$ from the AGN, and fainter emission from a radio lobe between this hot spot and the AGN. The width of the eastern radio lobe, and the more diffuse regions of the western jet, is about $3''$ ($\sim 25$ kpc), down to the $5\sigma$ level. 

The extended X-ray emission follows closely the radio jets to both the east and west. There is an X-ray hot spot toward the end of the eastern lobe, close to, but not exactly aligned with, the radio hot spot, with an offset of $\sim 0.7"$. Again, we estimate the relative X-ray and radio astrometry is accurate to $\sim 0.1"$, based on the strong AGN emission at both wavebands. We consider this X-ray -- radio hot spot offset in \citet{sobolewska}. 

\begin{figure}
\includegraphics[trim=-1in 4.5in 0in 1.5in, clip, width=\linewidth]{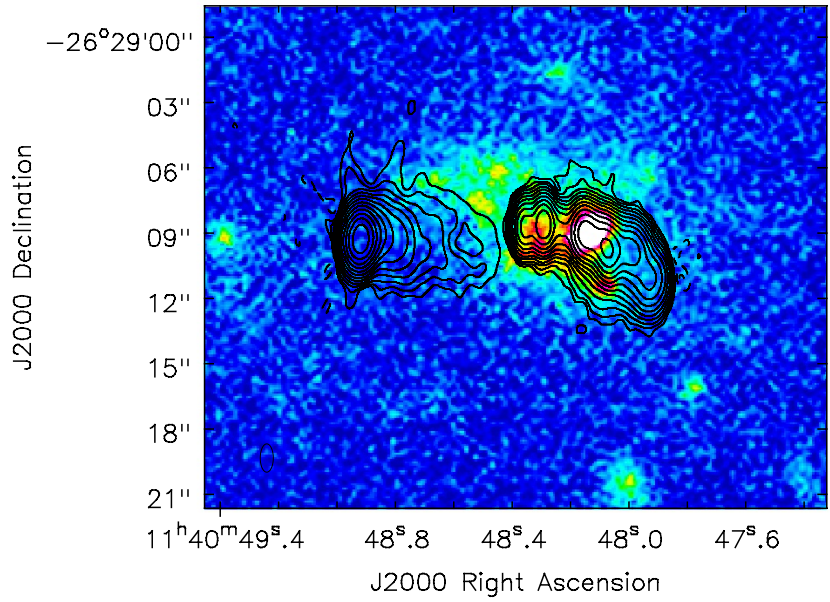}
\caption{Same contour image as in Figure 1. The color image is the ESO narrow band filter image, roughly centered on the Ly$\alpha$ emission line from the Spiderweb galaxy at $z = 2.16$ \citep{nonino}.
}
\label{fig:SLya}
\end{figure}

Figure~\ref{fig:SLya} shows the 3 GHz radio contours, and the ESO Ly$\alpha$ image \citep{nonino}. Extended Ly$\alpha$ emission is seen from the Spiderweb, which, at lower surface brightness, extends over more than 100 kpc, enveloping the protocluster \citep{md}. Herein, we focus on the brighter Ly$\alpha$ emission regions toward the center of the protocluster. Note that the Ly$\alpha$ emission does not peak at the location of the AGN, likely the result of dust obscuration toward the galaxy center. In general, obscuration will be an issue when interpreting UV measurements. 

There are two features of note with the comparison of the Ly$\alpha$ with the radio jets. First, on the western side, the radio jet deflects to the southwest just at a position of the brightest Ly$\alpha$ knot. This spatial coincidence suggests a physical association of the ambient thermal gas and the material within the radio jet, such as a strong shock where the fluids collide. This region will be discussed at length in a future paper, including the polarized radio emission \citep{anderson21}. 

The second Ly$\alpha$ -- radio (anti-) association is in the tail of the eastern lobe. There is a clear deficit of Ly$\alpha$ emission at the position of the diffuse radio lobe, and the Ly$\alpha$ emission follows the inner edge of the radio lobe. Again, this anti-coincidence suggests a physical association, in which the radio lobe somehow excludes the thermal gas, such as has been seen for eg. radio cavities in lower redshift X-ray clusters \citep{mcnamara05}. 

\begin{figure}
\includegraphics[trim=-1in 4.8in 0in 0.5in, clip, width=\linewidth]{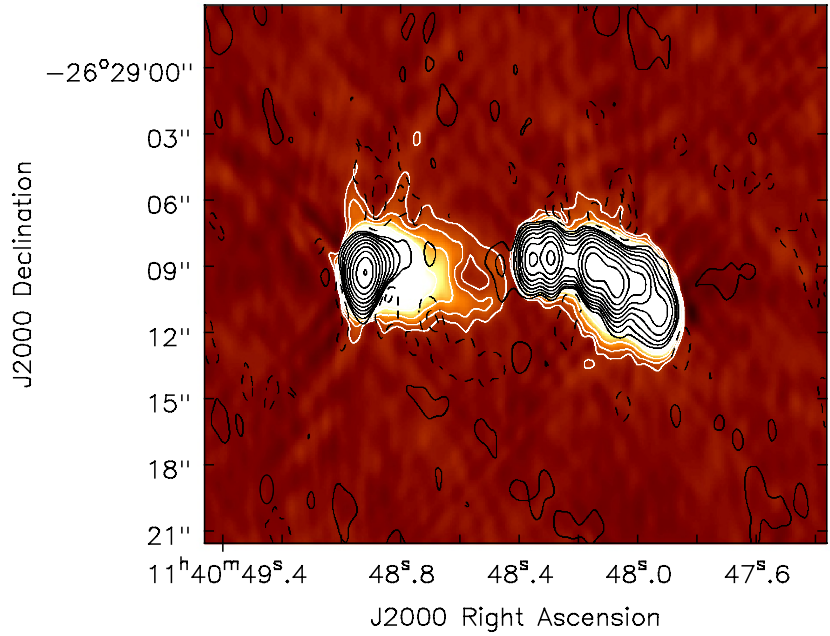}
\caption{The color scale is the 2 GHz to 4 GHz VLA image of the Spiderweb galaxy, as shown in Figure 1, saturated in intensity to show the low surface brightness emission. The white contours are the same as Figure 1 for the 2 GHz to 4 GHz image. The black contours are the 8 GHz to 12 GHz integrated emission, convolved to the same resolution as the S band image ($1.3''\times 0.6''$). The contour levels are a geometric progression in factor two, starting at 20 $\mu$Jy beam$^{-1}$. 
}
\label{fig:SX}
\end{figure}

Figure~\ref{fig:SX} shows the 10 GHz radio emission as black contours, plus a saturated color scale image of the 3 GHz radio emission, with the lower 3 GHz contours in white. The 10 GHz image has been convolved with the same Gaussian restoring beam as the 3 GHz image, and has an rms noise of $5~\mu$Jy beam$^{-1}$. This image shows the close association of the brighter radio knots all along the western jet, and for the eastern radio hot spot. However, the eastern radio lobe is undetected at 10 GHz down to very low surface brightness levels, implying a very steep radio spectrum, $\alpha < -2.5$.

\subsection{Spectra and Spectral Index}

\begin{figure}
\includegraphics[width=0.6\linewidth]{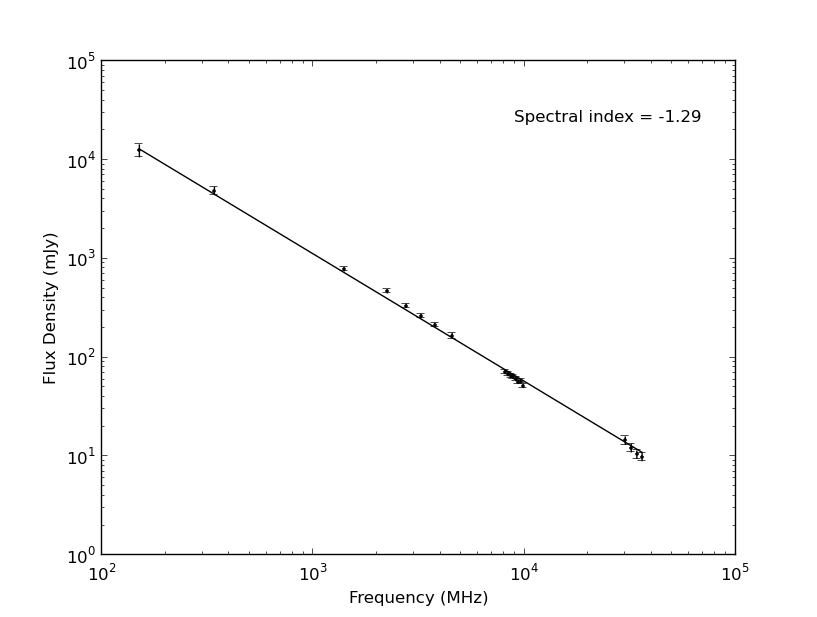}
\caption{Integrated radio spectrum of the Spiderweb radio galaxy. 
}
\label{fig:spec}
\end{figure}

Figure~\ref{fig:spec} shows the integrated radio spectrum of the Spiderweb from 150 MHz to 36 GHz. The data are roughly consistent with single power-law index of $\alpha = -1.3\pm 0.1$, across the full frequency range, where spectral index is defined as $S_\nu \propto \nu^\alpha$. There is evidence for modest steepening at high frequency, with the spectral index changing from $-1.15\pm 0.15$ between 150 MHz and 1450 MHz, to $-1.4\pm 0.1$ from 3 GHz to 32 GHz. The integrated spectrum mixes emission components across the source, including the core, flatter spectrum knots and steep spectrum lobes, and hence is of limited use for extrapolation of spectral properties in particular regions. 

\begin{figure}
\includegraphics[trim=-1in 5in 0in 0.5in, clip, width=\linewidth]{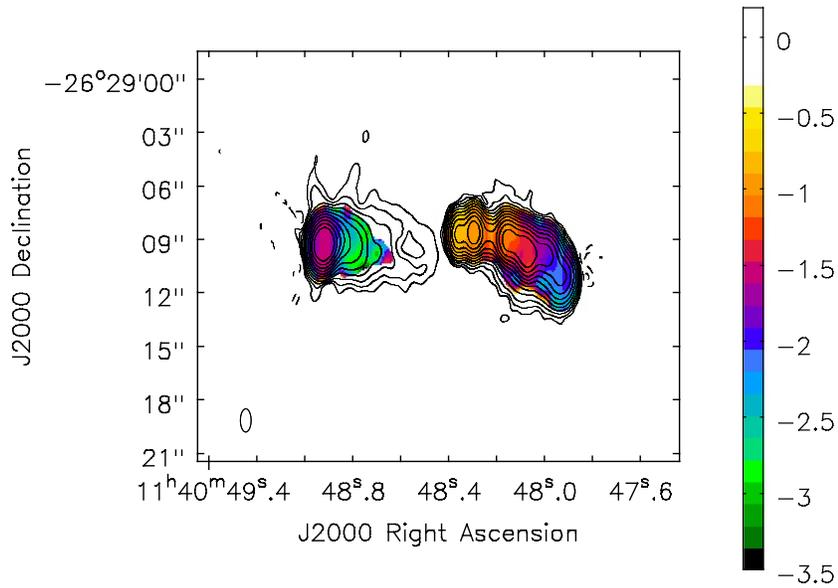}
\caption{The color scale is the spectral index across the 2 GHz to 4 GHz band, at the same resolution as shown in Figure 1, blanked at $8\sigma$. The contours are the same as Figure 1.
}
\label{fig:SIS}
\end{figure}

Figure~\ref{fig:SIS} shows the spectral index between images at 2.2 GHz and 3.8 GHz, along with the 3 GHz total intensity image. The radio core, and first western jet knot, have spectral indices $\le -1$, while the brighter knots further along the western jet, and the radio hot spot at the end of the eastern lobe, have steeper spectral indices $\sim -1.3\pm 0.2$. The outer jet to the west shows spectral steepening to $\alpha < -2$ with distance from the core, while the diffuse tail of the eastern lobe is very steep spectrum, with $\alpha \le -2.5$ in regions where it could be measured. For comparison, the spectral index study in \citet{pentericci97} shows steeper spectra in most regions between 4~GHz and 8~GHz, indicative of significant curvature (steepening), above 3~GHz. 

\begin{figure}
\includegraphics[trim=-1in 5in 0in 0.5in, clip, width=\linewidth]{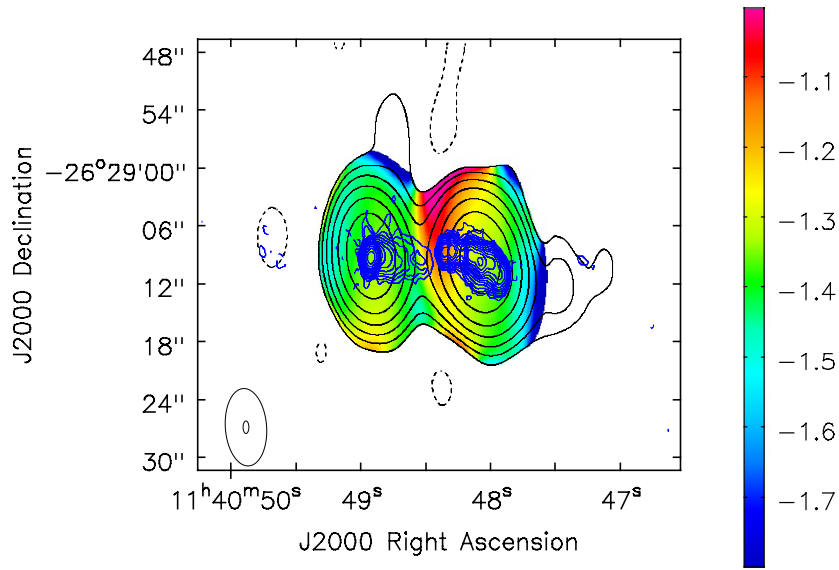}
\caption{The color scale is the spectral index between 340 MHz and 3 GHz, at the resolution of the 340 MHz image ($8.0''\times 4.3''$, major axis north-south). The black contours are the surface brightness at 340 MHz with geometric contour levels in factor two, starting at 20 mJy beam$^{-1}$. The blue contours are the same as in Figure 1. 
}
\label{fig:SISP}
\end{figure}

Figure~\ref{fig:SISP} shows the spectral index between 3 GHz and 340 GHz at lower spatial resolution. In this case, the resolution is adequate to only get two independent measurements: one for the western Jet, and one for the eastern hot spot plus radio lobe. Both are consistent with the integrated spectral index for the full source of $-1.3\pm 0.1$. 

\section{Analysis}

\subsection{Regions and Flux Densities}

Figure~\ref{fig:regions} shows the regions chosen for analysis. These regions were chosen to avoid emission from the bright AGN (radii $> 2.5''$ from the AGN), while capturing the outer eastern and western radio jet emission. We discuss the choice of regions in more depth below. Table \ref{table1} lists the 340 MHz and 3 GHz radio flux densities for the regions in columns 2 and 3. Column 4 lists the spectral indices. For the larger regions, the spectral index is between 340~MHz and 3~GHz. For the smaller regions (K3 and K4), the spectral index is between 2~GHz and 4~GHz. The errors include both off-source rms noise, plus gain uncertainties estimated during the flux bootstrap calibration process. For the west jet at 340 MHz, we subtracted the contribution of the AGN itself, and the innermost west jet knot, since these contribute to the low resolution 340 MHz flux density, but are outside the region used for the X-ray summation. The contributions (about 10\%), were estimated by extrapolating from 3 GHz using the measure spectral indices $\sim -1.2$ of these features (see Figure~\ref{fig:SIS}).

For the X-ray analysis of the jet emission, we have subtracted the background, and we have subtracted any residual contribution from the bright, compact AGN emission due to the far-out wings of the {\sl Chandra} point spread function. We have also subtracted possible diffuse thermal emission \citep{tozzi22}. These components contribute $\sim 25\%$ to the measured totals in the regions. The jet X-ray fluxes (in erg cm$^{-2}$ s$^{-1}$) in the soft band (0.5 keV to 2.0 keV), and hard band (2.0 keV to 10 keV), for the regions are listed in columns 5 and 6. Also listed are the power-law indices for the X-ray emission, derived from fitting power-law spectra jointly across the soft and hard X-ray bands (column 7). The X-ray spectral index in frequency, $\alpha$, relates to the X-ray photon index as: $\alpha$ = 1.0 - photon index. The X-ray flux errors are based on photon statistics, while the errors for the spectral indices are derived during the fitting process.  For the magnetic field and pressure analysis below, the dominant uncertainties arise from the assumptions in the model calculations, and we quote the model uncertainties explicitly where relevant. 

The X-ray frequency relates to photon energy, $E_{keV}$, as: $\nu_{x} = 2.44 \times 10^{17} \times E_{keV}$ Hz. We adopt an effective frequency, $\nu_x$, equal to the geometric mean of the low and high energy limits of a given {\sl Chandra} band (appropriate for a power-law spectrum of index $\sim -1$, consistent with the X-ray measurements). The corresponding values for the soft and hard bands are: $2.44\times 10^{17}$ Hz, and $1.11\times 10^{18}$ Hz, respectively. 

To derive X-ray flux densities required for the inverse Compton calculation below (in Jy = $10^{-23}$ erg s$^{-1}$ cm$^{-2}$ Hz$^{-1}$), we adopt a spectral index of $\alpha \sim -1.0$, for which the flux density becomes: $S_\nu = S_{band}/(\nu_x \times ln(\nu_{high}/\nu_{low}))$, where $S_{band}$ is the measured band-integrated flux (in erg cm$^{-2}$ s$^{-1}$), and $\nu_{high}$, $\nu_{low}$ are the high and low frequencies of the band. Again, $\nu_x$ is the geometric mean between $\nu_{high}$ and $\nu_{low}$. This equation is accurate to $\sim 15\%$ for $\alpha = -1.0 \pm 0.2$. The X-ray flux densities at the effective frequencies for the soft and hard {\sl Chandra} bands are given in columns 8 and 9 in Table \ref{table1}. We include the $15\%$ uncertainty in flux density due to spectral index uncertainty in the table. 

\subsection{X-ray Mechanisms}\label{sec:mech}

The very close association between X-ray and radio emission along both the eastern and western jets in the Spiderweb suggests physically related emission mechanisms. One possible mechanism to explain this spatial coincidence is X-ray synchrotron radiation from an extrapolation of the radio emitting relativistic electron population to very high energy. However, even if we assume no spectral steepening between radio and X-ray frequencies, extrapolating the radio flux densities into the X-ray under-predicts the observed X-ray flux densities by a factor $\sim 40$. Hence, the observed X-rays are much too bright to be an extrapolation of the radio synchrotron spectra to very high energy. 

A second mechanism to explain the relationship could be thermal emission from shocked gas surrounding the expanding radio source. This mechanism was favored by \citet{carilli02} based on shorter {\sl Chandra} observations, where the extended X-ray emission was detected at low significance and difficult to separate from the AGN itself. The deep X-ray image presented herein implies a substantially lower X-ray flux for the emission along the jet (the difference arising from a poor subtraction of the AGN in the \citet{carilli02} analysis). The new observations show that the X-ray emission is immediately coincident with the radio jets, and not along the edges of the radio source -- the latter being expected for thermal X-ray emission from a shocked cocoon \citep{snios}. 

Based on the morphology, flux, and spectral index (see Section~\ref{sec:freq}), we posit the most physically plausible mechanism to explain the Radio-X-ray association along the jet is inverse Compton up-scattering of the ambient photon field by the synchrotron emitting relativistic electrons. This mechanism has long been considered in the context of X-ray emission from cosmic radio sources \citep{hg79}, and it has been pointed out that up-scattering of the cosmic microwave background (CMB) photons could become a dominant X-ray emission mechanism in high redshift sources, due to the strong increase in the CMB energy density with redshift \citep{hodges-kluck,schwartz02, tavecchio00}. Numerous studies have now shown resolved X-ray emission from high redshift radio jets is best interpreted as inverse Compton emission \citep{scharf, cheung12, worrall20, marshall18, snios21, hodges-kluck,ighina}, and even in some sources without extended radio emission \citep{simionescu16, schwartz19}. 

\subsection{Magnetic Fields derived from CMB inverse Compton Emission}\label{sec:freq}

X-ray inverse Compton emission and the radio synchrotron emission arise from the same power-law distribution of relativistic electrons: the X-rays correspond to up-scattering of the ambient photon field, and the radio emission is, in a quantum mechanical sense, up-scattering of the virtual photons that comprise the static magnetic field. Hence, if one knows the energy density of the ambient photon field, then the X-ray emission effectively 'counts' the relativistic electrons, which, using the radio luminosity, then dictates the magnetic field strength \citep{longair}. 

An important consistency check is agreement between the X-ray and radio spectral indices, given the emission arises from the same power-law distribution of electrons. For all the regions analyzed herein, the spectral indices in the radio and X-ray agree, within the errors, with a value of $-1.2\pm 0.2$

Considering the ambient photon field, for high $z$ radio galaxies it has been shown that the ambient photons relevant for comparison of the observed radio synchrotron and X-ray inverse Compton emission are in the submm through far IR \citep{simionescu16, schwartz19,scharf}. While the Spiderweb is known to be luminous in the far IR \citep{seymour}, the dominant FIR emission is isolated to the inner $\le 1"$ radius \citep{gullberg}. \citet{gullberg} estimate an FIR luminosity of the host galaxy of $4.4\times 10^{12}$ L$_\odot$. For a centrally concentrated emitting region, the photon density behaves as $3L/(4 \pi r^2 c)$, where $L$ is the luminosity and $r$ the radius (eg. \citet{scharf}). The CMB energy density behaves as: $U_{CMB} = 4.2\times 10^{-13} \times (1+z)^4$ erg cm$^{-3}$. Comparing the CMB to the centrally-sourced FIR, one finds the CMB dominates the energy density beyond $\sim 20$~kpc, or $2.4''$, from the nucleus. Note that our high resolution images imply that the synchrotron photon energy density is below the CMB, even in the higher surface brightness regions.

For the preliminary analysis presented herein, we will focus on the outer regions of the source (radii $> 20$~kpc from the core), and adopt the CMB as the ambient photon field. A more complete treatment of the full extent of the source, including spectral contributions from various photon fields will be given in \citet{sobolewska}.  

The magnetic field calculation assuming up-scattering of the CMB was first presented by \citet{hg79}, and is given by:

\begin{center}
$ B_G^{(1-\alpha)} = 5.75 \times 10^{-17} \times (5.05\times 10^4)^{-\alpha}  \times (1+z)^{(3-\alpha)} \times [S_r \times \nu_r^{-\alpha}/S_x \times \nu_x^{-\alpha}] $
\end{center}

\noindent where $S_x$ and $S_r$ are the X-ray and radio flux densities at observed frequencies, $\nu_x$ and $\nu_r$ (all in matching units), and spectral index is defined as: $S = \nu^\alpha$. Again, $\alpha$ is dictated by the power-law relativistic electron energy distribution, and should be common to both the X-ray and radio emission. 

We examine four regions for which adoption of a spectral index of $-1.2\pm0.2$ in the X-ray and radio is reasonable. The regions are delineated in Figure \ref{fig:regions}. Two large regions capture the integrated X-ray emission and radio emission at 340 MHz, for the east and west jets beyond $\sim 2.5''$ from the AGN, respectively. These larger regions average over the source structure east and west of the nucleus in large boxes that contain all the emission, hence, the results are weighted toward the brighter radio hot spots and jet knots. We also consider spatially resolved measurements at 3~GHz of the two brightest knots beyond $2.5''$ from the AGN, for which the spectral index between 2~GHz and 4~GHz is consistent with $\alpha = -1.2 \pm 0.2$. 

Table~\ref{table1} lists the X-ray and radio flux densities for the four regions. Using the \citet{hg79} equation then leads to the magnetic fields listed in Table \ref{table2}. For the larger regions, we derive inverse Compton magnetic field strengths of $\sim 50~\mu$G. The implied magnetic energy density ($B^2/8\pi$ erg cm$^{-3}$, for $B$ in Gauss), is $\sim 1.0\times 10^{-10}$ erg cm$^{-3}$. For the two bright western jet knots, we obtain field values $\sim 70~\mu$G, and correspondingly a factor two higher magnetic energy densities. The field values derived from the Soft and Hard bands are consistent to within 20\%. An uncertainty in the spectral index of $0.2$ implies an uncertainty in the magnetic field of $\sim 20$\%.

\subsection{Minimum Energy Radio Synchrotron Calculations}\label{sec:minE}

Assuming a minimum energy configuration of relativistic particles and magnetic fields to estimate energy densities and magnetic fields in celestial synchrotron sources was first presented by \citet{burbidge56}. We adopt equations 1 and 2 in \citet{miley80}. Note that assuming energy equipartition between fields and relativistic particles, as opposed to a minimum energy configuration, leads to field values 8\% higher, and a total energy density within 1\% of the minimum \citep{readhead, carilli89}.

We assume the standard values of: electron/proton energy density = 1, filling factor = 1, a low frequency cutoff of 10 MHz, and a high frequency cutoff of 32 GHz \citep{miley80}. The calculation is relatively insensitive to the frequency cut-offs: raising the low frequency cutoff by a factor five increases the fields derived from the minimum energy assumption by 30\%, and the high frequency cut-off has an even smaller effect.

We have adopted a low frequency spectral index of $\alpha = -1.2$. Changing $\alpha$ by $\pm 0.2$ alters the derived fields by about 25\%, with steeper $\alpha$ leading to higher fields. In all cases, we assume a line-of-sight path length of 25 kpc ($\sim 3''$), corresponding to the width of the eastern lobe and western jet down to the $5\sigma$ surface brightness level. 

The field values and minimum energy densities are listed in Table \ref{table2}. The values derived are comparable to the early analysis in \citet{pentericci97}. The typical field values derived assuming minimum energy are about a factor two higher than those derived from the inverse Compton calculation. The implied minimum energy densities (summed fields and relativistic particles), are then about eight times higher than the magnetic energy density derived from the inverse Compton calculation. This factor eight arises from the fact that the minimum energy fields are $\sim 2\times$ higher, meaning a factor four higher magnetic energy density, and then there is another factor two when including the relativistic particles. 

\subsection{Pressures}

The minimum energy calculation for the synchrotron emission assumes that a minimum energy configuration is established by the relativistic particles and magnetic fields, implying equipartition of energy between fields and particles. There is no {\sl a priori} reason to assume such a minimum energy configuration is established. Hence, in the following analysis, we adopt the inverse Compton fields, and consider the implications for pressures within the radio source, in comparison to the ambient medium.  The relevant quantity is the total non-thermal pressure, meaning, the summed pressure in fields and relativistic particles. Hence, the synchrotron radio emission provides a critical constraint, since it must be explained in the context of the IC derived magnetic field strength, and hence dictates the total non-thermal pressure.

The synchrotron minimum energy density implies a minimum pressure, which, for an isotropic relativistic gas equals $1/3 ~\times$ energy density. Hence, the implied minimum pressure based on the synchrotron emissivity is $\sim 3\times 10^{-10}$ dyne cm$^{-2}$ (Table \ref{table2}). However, the inverse Compton magnetic field strengths are a factor of two lower than dictated by the minimum energy assumption. For fields lower than the minimum configuration, the total non-thermal energy density (and pressure), increases as $(B/B_{minE})^{3/2}$, in order to generate the observed radio emissivity. In this case, the pressure is dominated by the relativistic particles. For reference, for fields larger than $B_{minE}$, the total non-thermal pressure increases as $(B/B_{minE})^2$, as the total pressure becomes dominated by the field \citep{readhead, carilli89}. A field strength a factor two lower than minimum energy then implies a factor $\sim 3$ higher total non-thermal pressure $\sim 9\times 10^{-10}$ dyne cm$^{-2}$.

The first conclusions based on the X-ray and radio emission from the jets of the Spiderweb are that the non-thermal pressures within the radio source are larger than dictated by minimum energy by a factor three, and are dominated by relativistic particles, not magnetic fields. 

The Ly$\alpha$ emission provides a rough estimate of the thermal gas pressures in the Spiderweb protocluster. \citet{pentericci97} derive a typical density for the Ly$\alpha$ emitting knots in the Spiderweb of $\sim 40$ cm$^{-3}$. Assuming a temperature of $\sim 10^4$ K implies a thermal pressure of $5.5\times 10^{-11}$ dyne cm$^{-2}$. 

An important point to keep in mind is that the volume filling factor for the dense Ly$\alpha$ emitting gas is small ($\sim 10^{-5}$; \citep{mccarthy, pentericci98}). Hence, there must be a hot, intercloud medium confining the line emitting clouds. \citet{tozzi22} have nominally identified X-ray emission from this hot, diffuse medium in the Spiderweb, on the same spatial scale as the Ly$\alpha$ and radio jet emission, by subtracting the strong X-ray AGN, and by avoiding regions that may be contaminated by the X-ray jets. They estimate an average gas density and temperature in a sphere of radius 50~kpc of $n_e \sim 0.014\pm 0.0014$ cm$^{-3}$ and $T \sim 2.4_{-0.8}^{+0.5}\times 10^7$~K. The implied thermal pressure is then $4.6\pm 1.3 \times 10^{-11}$ dyne cm$^{-2}$. Encouragingly, this hot gas pressure is comparable to the pressure derived from the Ly$\alpha$ emission, and both imply an ambient thermal gas pressure an order of magnitude lower than the non-thermal pressures derived above for the bright radio emitting regions. A more detailed analysis of the cluster density radial profile is currently in progress.

The second conclusion therefore is that the high surface brightness radio emitting regions (hot spots and jet knots), are over-pressured relative to the ambient thermal gas in the cluster, by more than an order of magnitude. 

Assuming ram pressure confinement of the radio structures then implies a typical advance speed for the radio source $v_{ram} = (P_{nt}/\rho)^{1/2} \sim 2400$ km s$^{-1}$, where $\rho$ is the ambient mass density. Using the source maximum radius $\sim 60$ kpc implies a source age of $\sim 3\times 10^7$ years. 

\subsection{The Tail of the Eastern Lobe}

We re-emphasize that the calculations for the regions above are weighted to the higher brightness regions of the source, such as hot spots and jet knots, and hence to regions of higher pressure. As a final calculation, we consider the synchrotron minimum energy density in the very low surface brightness, steep spectrum tail of the eastern radio lobe (but still beyond 20~kpc from the core). This is the region that shows the 'cavity' in the Ly$\alpha$ emission coincident with the radio lobe. 

The difficulty in calculating the pressure in the diffuse radio lobe is, again, that spatially resolving observations are only available at 3 GHz, and that the spectrum at this high frequency is seen to be very steep, $\sim -2.5$, but must flatten to lower frequency to explain the integrated radio spectrum of the source. In this case, we use the typical surface brightness in the tail of the radio lobe of $\sim 50~\mu$Jy beam$^{-1}$ at 3 GHz, where the beam is $1.3''\times 0.6''$. We extrapolate the 3~GHz surface brightness to 1 GHz using the steep spectral index, implying a surface brightness at 1~GHz of $\sim 780~\mu$Jy beam$^{-1}$. We then calculate the synchrotron minimum energy density using this 1 GHz surface brightness, and the lower frequency spectral index, $\alpha = -1.2$, as above. This calculation leads to a magnetic field strength of $37~\mu$G, a minimum non-thermal energy density of $1.3\times 10^{-10}$ erg cm$^{-3}$, and a minimum non-thermal pressure of $4\times 10^{-11}$ dyne cm$^{-2}$. Decreasing the turn-over frequency from 1 GHz to 0.5 GHz increases the minimum non-thermal pressure by 60\%. 

Hence, if the diffuse tails of the radio lobes have a spectral flattening between 0.5 GHz and 1 GHz, and if the fields and particles have been able to reach a minimum energy configuration in these potentially more relaxed and older regions, then the non-thermal radio lobe pressure is comparable to the thermal pressures of the ambient medium. 

\section{Discussion}\label{sec:5}

We comment on one curiosity and a general question. Considering the radio source morphology, the eastern jet of the Spiderweb shows a radio hot spot at the extremity of the lobe, with clear spectral steepening from the hot spot back toward the AGN. Conversely, the western jet shows a gradual decline in surface brightness, and spectral steepening with distance from the AGN. The eastern morphology, in both intensity and spectrum, is characteristic of high power radio Fanaroff-Riley Class II radio galaxies ('edge-brightened', typically with $P_{178MHz} > 10^{35}$ erg s$^{-1}$ Hz$^{-1}$), while the western jet is typical of lower power FRI ('edge-darkened'), radio jets \citep{fr}. However, the Spiderweb is one of the most luminous radio galaxies known, with $P_{178MHz} \sim 4.6\times 10^{36}$ erg s$^{-1}$ Hz$^{-1}$, placing it several orders of magnitude above the canonical FRII - FRI break. Why then the dichotomy in both morphology and spectral index between lobes? 

The Spiderweb is a high-redshift example of a hybrid morphology radio source (or HyMoRS), which are a known but rare and poorly understood class of radio galaxy, first described by \citet{gopal00}, and for which correlated differences in morphology and spectral index between lobes have previously been observed \citep{gawronski06, deGasperin17}. Their nature is not well constrained: They may arise because of the physical attributes of the source (e.g. jet power) or its environment \citep{kapinska17}, or because of observational or orientation effects \citep{harwood20}, or by some combination of these. 

In the case of the Spiderweb system, a potential explanation is seen in the Ly$\alpha$ emission and galaxy density in the center of the cluster. The thermal gas and galaxies are asymmetrically distributed in the cluster, with a higher density toward the west than the east. It is possible that this density inhomogeneity, in a dynamically active and young protocluster, leads to more disturbed jet advance toward the west, while the eastern jet remains relatively unperturbed until it reaches the terminal jet shock (hot spot). The perturbed nature of the western jet is supported by the bright Ly$\alpha$ knot at the jet deflection point, discussed above. 

We note the very steep spectra for the diffuse emitting regions of the Spiderweb, such as in the eastern lobe, and the end of the western jet, where $\alpha < -2$ (see Figure~\ref{fig:SX} and Figure~\ref{fig:SIS}).  This may, in part, be due to increasing inverse Compton energy losses off the CMB for the relativistic electrons. Indeed, this phenomenon has been invoked to explain the correlation between redshift and ultra-steep spectrum radio galaxies (for a review, see \citet{md}). The VLA snapshot survey of \citet{pentericci00} shows evidence that the majority of $z > 2$ powerful radio galaxies have lobe regions with very steep spectra ($\alpha < -2$ between 5~GHz and 8~GHz), although the snap-shot sensitivity of the pre-upgraded VLA was inadequate to properly image these regions in detail. 

A general question is: given the rapid rise with redshift of the energy density in the CMB, at what redshift will cooling of the relativistic electrons be dominated by inverse Compton emission over synchrotron radiation? The calculation is straight-forward, since the ratio of radio synchrotron to X-ray luminosity is given simply by: $L_r/L_x = U_B/U_{CMB}$ \citep{schwartz02, ct02}. The magnetic energy density in a radio source is: $U_B \sim 1\times 10^{-10} (B/50\mu G)^2$ erg cm$^{-3}$, while the CMB energy density is: $U_{CMB} = 4.2\times 10^{-13} \times (1+z)^4$ erg cm$^{-3}$. Hence, $L_x > L_r$ at: $(1 + z) > 3.9\times (B/50\mu G)^{1/2}$. For typical radio jet magnetic fields of 50~$\mu$G, inverse Compton dominates over radio synchrotron losses at $z > 2.9$. A recent analysis of 11 radio galaxies from $z = 1.3$ to 4.3 shows that, for the diffuse emission, the X-ray and radio data are consistent with both Inverse Compton scattering of the CMB, and with equipartition magnetic fields \citep{hodges-kluck}.

\section{Summary}

Using the observed inverse Compton and radio synchrotron radiation, we derive typical magnetic field strengths in the jets of the Spiderweb $\sim 50~\mu$G to $70 \mu$G, in regions of the source beyond 20~kpc from the core. These are regionally averaged values, derived over areas of significant radio structure, hence represent a brightness weighted mean, ie. weighted toward knots and hot spots, corresponding to regions of particle acceleration due to shocks or regions of high compression. 

The inverse Compton fields are a factor two lower than those derived assuming a minimum energy ($\sim$ equipartition), configuration for the synchrotron emitting relativistic particles and fields. The implied non-thermal pressure (sum of relativistic particles and fields), implied by the combination of inverse Compton X-rays and radio synchrotron radiation, is $\sim 9\times 10^{-10}$ dyne cm$^{-2}$. This non-thermal pressure is a factor three higher than the pressure derived by assuming minimum energy conditions, and the pressure is dominated by relativistic particles. Comparing the typical non-thermal pressure in the radio source with the thermal pressure in the ambient cluster gas derived from Ly$\alpha$ emission and from possible diffuse thermal X-ray emission, implies a factor 15 higher pressures in the brighter radio emitting regions than in the ambient thermal gas. Assuming ram pressure confinement of the radio source by an hypothesized hot, diffuse cluster gas, implies an advance speed $\sim 2400$ km s$^{-1}$, and a radio source lifetime of $\sim 3\times 10^7$ years.

We see two regions where there is a clear spatial correlation between radio structures and the Ly$\alpha$ emitting gas. One region is the brightest Ly$\alpha$ knot in the western jet coincident with the position where the radio jet deflects toward the southwest. This deflection suggests a strong shock due to the impact of the radio jet on an ambient density enhancement. We will discuss this region in detail in our polarization analysis of the radio emission \citep{anderson21}. 

The second spatial association of the Ly$\alpha$ and radio emission is seen in the tail of the eastern lobe, where the radio lobe appears evacuated of Ly$\alpha$ emission, while the line emission outer edge (away from the AGN), aligns closely with the inner edge of the radio lobe. This anti-correlated morphology is reminiscent of cavities in X-ray clusters due to expanding radio sources \citep{mcnamara05}. A calculation of the minimum pressure in the non-thermal gas in the faintest radio emitting regions, based on 3 GHz surface brightness, and assuming a spectral flattening between 0.5 GHz and 1.0 GHz, leads to approximate pressure equilibrium between the minimum non-thermal pressures and the ambient thermal cluster gas. In the context of the standard model for powerful radio galaxies, in which the lobes represent the older, more relaxed, populations of relativistic particles and fields left behind by the expanding radio source, it seems physically plausible that a minimum energy configuration for fields and particles would be more likely in the relaxed lobes. Spatially resolving observations at low frequency would help to test this hypothesis. 

We show that the Spiderweb radio source is a high redshift example of a 'Hybrid Morphology' radio source, with the western jet have the morphology in total intensity and spectral index of an FRI-class radio galaxy (edge-darkened), and the eastern region having the morphology of an FRII-class radio galaxy (edge-brightened) \citep{gopal00}. The origin of such HyMoRS remains an area of active research. In the case of the Spiderweb, a clue to the origin may be seen in the asymmetric gas environment, for which the Ly$\alpha$ emission is predominantly west of the nucleus, potentially leading to higher disruption of the advancing radio jet. 

\acknowledgments
This program employed VLA data from project 19A-024. The National Radio Astronomy Observatory is a facility of the National Science Foundation operated under cooperative agreement by Associated Universities, Inc.. CLC acknowledges support through CXC grant G09-2-1103X. SB acknowledges financial support from the INFN INDARK grant and the agreement ASI-INAF n.2017-14-H.0. MP, AS and LDM are supported by the ERCStG ‘ClustersXCosmo’ grant agreement 716762. AS is also supported by the FARE-MIUR grant 'ClustersXEuclid' R165SBKTMA, and by INFN InDark Grant.

\clearpage
\newpage

\begin{landscape}
\begin{table}
\centering
\footnotesize
\caption{Observed Properties}
\begin{tabular}{lccccccccc} 
\hline
 Region &  S$_{340 MHz}$ & S$_{3 GHz}$ & $\alpha_{0.34}^{3}$ or $\alpha_{2}^{4}$ & S$_{soft}$ & S$_{hard}$ & $\alpha_{soft}^{hard}$ & S$_{2.4e17 Hz}$ & S$_{1.1e18 Hz}$ &  Area  \\
 ~ & Jy & mJy & ~ & $10^{-15}$ erg~cm$^{-2}$~s$^{-1}$ &  $10^{-15}$ erg~cm$^{-2}$~s$^{-1}$ & ~ &  $10^{-11}$ Jy & $10^{-11}$ Jy & arcsec$^2$ \\
 \hline
West Jet & $2.6\pm0.4$ & $170\pm 12$ & $-1.2\pm 0.1$ & $1.2\pm 0.13$  & $1.4\pm 0.14$  & $-1.0 \pm 0.2$ & $35\pm 6$  & $7.9\pm 1.4$  & 27 \\
East Jet & $1.8\pm 0.3$ & $111\pm 8$ & $-1.3\pm 0.1$ & $1.5\pm 0.14$  & $1.1\pm 0.12$  & $-1.3\pm 0.2$ & $46\pm 8$  & $6.4\pm 1.1$  &  30 \\
K3 & -- & $51\pm 4$ & $-1.2\pm0.2$ & $0.22\pm 0.05$ & $0.16\pm 0.04$ & $-1.3 \pm 0.2$ & $6.5\pm 1.8$ & $0.92\pm 0.24$ & 2.4 \\
K4 & -- & $86\pm 6$ & $-1.4\pm0.2$ & $0.33\pm 0.6$ & $0.41\pm 0.8$ & $-1.0 \pm 0.2$ & $9.8\pm 2.4$ & $2.3\pm 0.6$ & 5.1 \\
\hline
\vspace{0.1cm}
\end{tabular}
\label{table1}
\end{table}

\begin{table}
\centering
\footnotesize
\caption{Magnetic Fields and Energy Densities}
\begin{tabular}{lcccccc} 
\hline
 Region &  B$_{IC,soft}$ & U$_{BIC,soft}$ & B$_{IC,hard}$ & U$_{BIC,hard}$ & B$_{minE}$ & U$_{minE}$ \\
 ~ & $\mu$G & erg cm$^{-3}$ & $\mu$G & erg cm$^{-3}$ & $\mu$G & erg cm$^{-3}$ \\
 \hline
West Jet & 60 & $1.4 \times 10^{-10}$  & 54 & $1.2 \times 10^{-10}$ & 103 & $9.9 \times 10^{-10}$ \\
East Jet & 45 & $8.1 \times 10^{-11}$ & 50 & $1.0 \times 10^{-10}$ & 95 & $8.4 \times 10^{-10}$  \\
West Jet K3 & 68 & $1.8 \times 10^{-10}$  & 76 & $2.3 \times 10^{-10}$ & 125 & $1.4 \times 10^{-9}$ \\
West Jet K4 & 72 & $2.0 \times 10^{-10}$  & 63 & $1.6 \times 10^{-10}$ & 125 & $1.4 \times 10^{-9}$ \\
\hline
\vspace{0.1cm}
\end{tabular}
\label{table2}
\end{table}

\end{landscape}

\clearpage
\newpage

\end{document}